# Nonsymmetric Dependence Measures: the Discrete Case

*Hui Li*[1]

Following our previous work on copula-based nonsymmetric dependence measures, we introduce similar measures for discrete random variables. The measures cover the range between two extremes: independence and complete dependence, which take minimum value exactly on independence and take maximum value exactly on complete dependence. We find that the ∗ product on copulas in the continuous case reduces to matrix product of transition matrices in the discrete case and we use it to prove the DPI condition. The measures can also be extended to detect dependence between groups of discrete random variables or conditional dependence. Unlike the continuous case, one drawback is that the value of the measures depends on marginal distributions.

1.  Introduction

In two previous papers (Li, 2015a and 2015b), we proposed copula-based nonsymmetric bivariate and multivariate dependence measures that capture both independence and complete dependence between random variables. The measures are nonsymmetric in order to accurately detect complete dependence or functional relationship of one variable on another variable. This means that they take minimum value exactly on independence and take maximum value exactly on complete dependence of one variable on another but not the opposite way. Here for completeness we will extend our work to introduce measures of dependence for discrete random variables. The continuous case can be made nonparametric or independent of marginal distributions through copula, which is no longer true for the discrete case as the value of the dependence measures will be dependent on the marginal distributions.

In the discrete case, a well-known quantity that has often been used as dependence measure is Shannon's mutual information (Shannon and Weaver, 1949). Its limitation in the continuous case has been discussed in Li (2015a). The issue is that it becomes infinity whenever the copula has singular component, even if there is no functional relationship. We will use an example to show that, in the discrete case, it also has the similar issue where the new nonsymmetric dependence measures give more meaningful results.

As conditional mutual information is also used in causality detection for time series analysis (Hlaváčková-Schindler, Paluš, Vejmelka and Bhattacharya, 2007), we extend the new dependence measures to the conditional case so they can be useful in that area.

The structure of the current paper is as follows. Section 2 presents the bivariate nonsymmetric dependence measures for discrete random variables including both distance and entropy forms. Again we prove the Data Processing Inequality (DPI) for the measures with the help of transition matrix on Markov chain. We then compare the measure in distance form with Shannon's mutual information through an example. Section 3 presents the nonsymmetric multivariate dependence measures between two groups of discrete random variables. Section 4 presents a conditional nonsymmetric dependence measure, which can be used to replace conditional mutual information for causality detection. Section 5 concludes the paper.

## 2. Bivariate discrete dependence measures

As suggested in Li (2015a, 2015b), a nonsymmetric dependence measure can be constructed from the distance between conditional cumulative distribution and unconditional cumulative distribution.

Let $X, Y$ be discrete random variables with values in $\{x_i\}_{i=1}^m$, $\{y_j\}_{j=1}^n$ respectively. Let $P(i) = P(X = x_i)$, $P(j) = P(Y = y_j)$ be the marginal probabilities and $P(i,j) = P(X = x_i, Y = y_j)$ be the joint probability. Let $P(j|i) = P(Y = y_j|X = x_i) = P(X = x_i, Y = y_j)/P(j)$ be the conditional probability, $F(j) = \sum_{j'=1}^{j} P(j')$ and $F(j|i) = \sum_{j'=1}^{j} P(j'|i)$ be the unconditional and conditional cumulative probability.

The Shannon's mutual information is defined as

$$I(X, Y) = \sum_{i,j=1}^{m,n} P(i,j) \cdot \log_2 \frac{P(i,j)}{P(i) \cdot P(j)} \tag{1}$$

which is symmetric as $I(X, Y) = I(Y, X)$.

Let us define our new nonsymmetric measure of dependence of $Y$ on $X$ as

$$\tau(X, Y)^2 = 6 \sum_{i,j=1}^{m,n} [F(j|i) - F(j)]^2 \cdot P(i) \cdot P(j) \tag{2}$$

Obviously $\tau(X, Y) \geq 0$ with the equal sign holds if and only if $F(j|i) = F(j)$ or $Y$ is independent of $X$. On the other hand,

$$\tau(X, Y)^2 = 6 \sum_{i,j=1}^{m,n} F(j|i)^2 \cdot P(i) \cdot P(j) - 6 \sum_{j=1}^{n} F(j)^2 \cdot P(j)$$

$$\leq 6 \sum_{i,j=1}^{m,n} F(j|i) \cdot P(i) \cdot P(j) - 6 \sum_{j=1}^{n} F(j)^2 \cdot P(j)$$

$$= 6\sum_{j=1}^{n} F(j) \cdot P(j) - 6\sum_{j=1}^{n} F(j)^2 \cdot P(j)$$

$$= 6\sum_{j=1}^{n} F(j) \cdot (1 - F(j)) \cdot P(j) \tag{3}$$

where we have used $0 \leq F(j|i) \leq 1$ and

$$\sum_{i=1}^{m} F(j|i) \cdot P(i) = F(j). \tag{4}$$

The equal sign in Equation (3) holds if and only if $F(j|i) = 0$ or $1$ for all $1 \leq i \leq m, 1 \leq j \leq n$. Since $F(j|i)$ is a non-decreasing function in $j$ and $F(n|i) = 1$, this means that, at a certain value of $j$, $F(j|i)$ jumps from 0 to 1 for each $i$. So $Y$ is a function of $X$. The maximum value of $\tau(X,Y)^2$ can be viewed as a discrete approximation to the continuous integral $6\int_0^1 x(1-x)dx = 1$, which is the constant maximum value for the continuous bivariate case (Li, 2015a). However, in the discrete case, this maximum value depends on the distribution of $Y$, although it is in a tight range $(0, \frac{3}{2})$ as $0 \leq F(j) \cdot (1 - F(j)) \leq \frac{1}{4}$. This kind of phenomenon has been noticed before in Giannerini, Maasoumi, and Dagum (2015), where the value of the discrete version of the Bhattacharya-Hellinger-Matusita distance depends on the marginal distribution. In general, this also happens to concordance measures which have well-defined range in the continuous case, but depend on marginal distributions in the discrete case, see Genest and Nešlehová (2007).

The benefit of the new measure comparing with traditional measures such as mutual information is that it characterizes both independence and complete dependence of a functional relationship as will be discussed in an example later in the section. Traditional measures can take maximum value even when there is no functional relationship.

Next we introduce the entropy form of the dependence measure following Rényi (1961) and Li (2015a):

$$R_\alpha(X,Y) = \frac{1}{\alpha - 1} \log \left( \sum_{i,j=1}^{m,n} \left( \frac{F(j|i)}{F(j)} \right)^\alpha \cdot P(i) \cdot P(j) \right) \quad \text{for } 0 < \alpha < 2 \tag{5}$$

We have

$$R_\alpha(X,Y) = \frac{1}{\alpha - 1} \log \left( \sum_{i,j=1}^{m,n} \frac{F(j|i)}{F(j)} \cdot \left( \frac{F(j)}{F(j|i)} \right)^{1-\alpha} \cdot P(i) \cdot P(j) \right)$$

$$\geq \frac{1}{\alpha - 1} \log \left( \sum_{i,j=1}^{m,n} \frac{F(j|i)}{F(j)} \cdot \frac{F(j)}{F(j|i)} \cdot P(i) \cdot P(j) \right)^{1-\alpha} = 0 \tag{6}$$

where we have used Jensen's inequality and the following

$$\sum_{i,j}^{m,n} \frac{F(j|i)}{F(j)} \cdot P(i) \cdot P(j) = 1 \tag{7}$$

The lower bound holds when $F(j|i) = F(j)$, or $X, Y$ are independent. Meanwhile, as $0 \leq F(j|i) \leq 1$, $F(j|i)^\alpha \leq F(j|i)$ when $1 < \alpha < 2$ and $F(j|i)^\alpha \geq F(j|i)$ when $0 < \alpha < 1$. So we have

$$R_\alpha(X,Y) \leq \frac{1}{\alpha-1} \log\left(\sum_{i,j=1}^{m,n} \frac{F(j|i)}{F(j)^\alpha} \cdot P(i) \cdot P(j)\right) = \frac{1}{\alpha-1} \log\left(\sum_{j=1}^{n} F(j)^{1-\alpha} \cdot P(j)\right) \quad (8)$$

The upper bound holds when $F(j|i) = 0 \text{ or } 1$ for all $i, j$. As $F(j|i)$ is non-decreasing in $j$, this means $F(j|i)$ jumps from 0 to 1 at certain $j$ for each $i$. This only happens when $P(j|i) = 1$ for exactly one $j$ for each $i$, or $Y$ is a function of $X$. Again, the value of the upper bound is a discrete approximation of the continuous case $\frac{1}{\alpha-1} \log\left(\int_0^1 v^{1-\alpha} dv\right) = -\frac{\log(2-\alpha)}{\alpha-1}$, see Li (2015a), Equation (A3). When $\alpha \geq 2$, the measure may not be bounded.

Another entropy form follows Tsallis (1988) and Li (2015a),

$$\Delta_\alpha(X,Y) = \frac{1}{\alpha-1}\left(\sum_{i,j=1}^{m,n} \left(\frac{F(j|i)}{F(j)}\right)^\alpha \cdot P(i) \cdot P(j) - 1\right) \qquad \text{for } 0 < \alpha < 2 \quad (9)$$

In the limit $\alpha \to 1$, $R_\alpha(X,Y)$ and $\Delta_\alpha(X,Y)$ both reduce to

$$R(X,Y) = \sum_{i,j=1}^{m,n} \frac{F(j|i)}{F(j)} \log \frac{F(j|i)}{F(j)} \cdot P(i) \cdot P(j) \quad (10)$$

which has lower bound 0 and upper bound $-\sum_{j=1}^{n} \log F(j) \cdot P(j)$. The upper bound is a discrete approximation to the integral $-\int_0^1 \log x \cdot dx = 1$.

Let $Z$ be another discrete random variable with values in $\{z_k\}_{k=1}^{l}$. If $X$ and $Z$ are independent conditional on $Y$, then we have $P(i,k|j) = P(i|j) \cdot P(k|j)$ and

$$P(i,k) = \sum_{j=1}^{n} P(i,k|j) \cdot P(j) = \sum_{j=1}^{n} P(i|j) \cdot P(k|j) \cdot P(j) = \sum_{j=1}^{n} P(i,j) \cdot P(k|j)$$

such that, if both sides are divided by $P(i)$,

$$P(k|i) = \sum_{j=1}^{n} P(k|j) \cdot P(j|i) \quad (11)$$

This is the discrete form of the $*$ product for copulas as described in (Li, 2015a), Equations (14) and (29). If we identify $P(j|i)$ as the transition matrix $M_{XY}$ from $X$ to $Y$, then Equation (11) becomes matrix multiplication,

$$M_{XZ} = M_{XY} \cdot M_{YZ} \quad (12)$$

Equation (11) also implies

$$F(k|i) = \sum_{j=1}^{n} F(k|j) \cdot P(j|i) \quad (13)$$

The transition matrix is also called singly stochastic matrix. The set of all singly stochastic matrix forms a Markov algebra as discussed in Darsow, Nguyen and Olsen (1992), whose work also motivated the research on nonsymmetric dependence measures for the continuous case.

A general nonsymmetric dependence measure is of the form,

$$\tau(X,Y) = \sum_{i,j=1}^{m,n} \varphi(F(j|i) - F(j)) \cdot P(i) \cdot P(j) \tag{14}$$

where $\varphi$ is a convex function and is not explicitly dependent on $i$.

PROPOSITION 2.1 If $X, Y, Z$ form a Markov chain, then

$$\tau(X,Z) \leq \tau(Y,Z) \tag{15}$$

Proof:

$$\begin{aligned}
\tau(X,Z) &= \sum_{i,k=1}^{m,l} \varphi\big(F(k|i) - F(k)\big) \cdot P(i) \cdot P(k) \\
&= \sum_{i,k=1}^{m,l} \varphi\left(\sum_{j=1}^{n} F(k|j) \cdot P(j|i) - F(k)\right) \cdot P(i) \cdot P(k) \\
&= \sum_{i,k=1}^{m,l} \varphi\big(\sum_{j=1}^{n}(F(k|j) - F(k)) \cdot P(j|i)\big) \cdot P(i) \cdot P(k) \\
&\leq \sum_{i,k=1}^{m,l} \left(\sum_{j=1}^{n} \varphi\big(F(k|j) - F(k)\big) \cdot P(j|i)\right) \cdot P(i) \cdot P(k) \\
&= \sum_{j,k=1}^{n,l} \varphi\big(F(k|j) - F(k)\big) \cdot P(j) \cdot P(k) \\
&= \tau(Y,Z) \tag{16}
\end{aligned}$$

where we used the property that $\sum_{j=1}^{n} P(j|i) = 1$ irrespective of $i$ and Jensen's inequality. As $Z$, $Y$, $X$ also form a Markov chain, we have $\tau(Z,X) \leq \tau(Y,X)$. The equal sign in Equation (16) holds when $F(k|j) - F(k)$ is constant relative to the distribution $P(j|i)$ on $j$ for any $k, i$. □

This is the discrete form of the Data Processing Inequality for nonsymmetric dependence measure, compare Li (2015a). It means that $Z$ is less dependent on $X$ than on $Y$ as $Y$ is closer to $Z$ on the Markov chain. Data Processing Inequality is known to be an important property of Shannon's mutual information (Cover and Thomas, 1991), which formalizes the idea that information is generally lost when transmitted through a noisy channel.

PROPOSITION 2.2. If $f$ is a bijective transformation on $X$, then $\tau(f(X), Y) = \tau(X, Y)$.

Proof: As $f$ is bijective, $X, f(X), Y$ form a Markov chain. So $\tau(X,Y) \leq \tau(f(X), Y)$. On the other hand, $f(X), X, Y$ also form a Markov chain, which leads to $\tau(f(X), Y) \leq \tau(X, Y)$. □

If $f$ is a permutation on values of $X$, then the dependence measure will not change, which may not be true for $Y$. It is easy to see that Equation (2) is invariant under permutation on $i$.

Another issue with the discrete nonsymmetric dependence measure is that it depends on the order of the discrete values of $Y$ in the calculation of cumulative conditional probability, which can be arbitrary. Thus it is not invariant under permutations on $j$. The default order we choose is by the real values of the random variable, which is consistent with the continuous case. However, the difference between different orderings becomes less as the number of elements increases and the distribution is smoother.

Next we discuss an example that shows the problem with Shannon's mutual information where the new dependence measure actually proves useful.

Let $Y$ be a random variable taking values in $\{\frac{1}{4n}, \frac{2}{4n}, \cdots, \frac{4n-1}{4n}, 1\} - \{\frac{1}{4}, \frac{1}{2}, \frac{3}{4}, 1\}$ with equal probability $\frac{1}{4n-4}$ for $n > 1$. Let $X = \cos(2\pi Y)$ and $Z = \sin(2\pi Y)$ such that $X^2 + Z^2 = 1$ is always true. We take out the 4 points $\{\frac{1}{4}, \frac{1}{2}, \frac{3}{4}, 1\}$ in $Y$ to avoid degeneracy in $X$ or $Z$. Using the fact that $X$ and $Z$ are each uniformly distributed with $2n - 2$ values, and $(X, Z)$ is uniformly distributed on $4n - 4$ value pairs, it is easy to calculate the symmetric mutual information for each pair of random variables:

$$I(X,Y) = I(Y,Z) = \sum_{i=1}^{4n-4} \frac{1}{4n-4} \log_2 \frac{\frac{1}{4n-4}}{\frac{1}{2n-2}\cdot\frac{1}{4n-4}} = \log_2(2n-2) \tag{17}$$

$$I(X,Z) = \sum_{i=1}^{4n-4} \frac{1}{4n-4} \log_2 \frac{\frac{1}{4n-4}}{\frac{1}{2n-2}\cdot\frac{1}{2n-2}} = \log_2(2n-2) - 1 \tag{18}$$

In the limit $n \to \infty$, $I(X,Y) = I(Y,Z) = I(X,Z) = \infty$. So, although $X$ is not a function of $Z$ or vice versa, $I(X,Z)$ is only 1 bit less than $I(X,Y)$ or $I(Y,Z)$, and becomes large for large $n$. That extra bit is the extra information needed to code $X$ given $Z$ or vice versa. If we convert the mutual information into the information coefficient of correlation (Linfoot, 1957) defined as $I_c(X,Y) = \sqrt{1 - e^{-2I(X,Y)}}$, then all three pairs give the value 1 when $n \to \infty$.

We note that the discrete version of the symmetric Bhattacharya-Hellinger-Matusita distance

$$S_\rho = \frac{1}{2}\sum_{i,j=1}^{m,n}\left(1 - \sqrt{\frac{P(i)\cdot P(j)}{P(i,j)}}\right)^2 \cdot P(i,j) = 1 - \sum_{i,j=1}^{m,n}\sqrt{P(i)\cdot P(j)\cdot P(i,j)} \tag{19}$$

also has the same issue, although it has many desirable properties (Granger, Maasoumi and Racine, 2004). It is related to the symmetric version of the Tsallis entropy with $\alpha = \frac{1}{2}$. Specifically, we have

$$S_\rho(X,Y) = S_\rho(Y,Z) = 1 - \sum_{i=1}^{4n-4} \sqrt{\frac{1}{4n-4} \cdot \frac{1}{2n-2} \cdot \frac{1}{4n-4}} = 1 - \frac{1}{\sqrt{2n-2}} \qquad (20)$$

$$S_\rho(X,Z) = 1 - \sum_{i=1}^{4n-4} \sqrt{\frac{1}{4n-4} \cdot \frac{1}{2n-2} \cdot \frac{1}{2n-2}} = 1 - \frac{1}{\sqrt{n-1}} \qquad (21)$$

In the limit $n \to \infty$, $S_\rho(X,Y) = S_\rho(Y,Z) = S_\rho(X,Z) = 1$, although there is no functional relationship between $X, Z$ and $Y$ is not a function of $X$ or $Z$. In the continuous case, this issue still persists.

For our new nonsymmetric dependence measure defined in Equation (2), we order the values of the random variables $X, Y$ and $Z$ by their real values for calculating cumulative distribution. For the dependence of $X$ or $Z$ on $Y$, as they are functions of $Y$, it is straight forward to use the upper bound in Equation (3),

$$\tau(Y,X)^2 = \tau(Y,Z)^2 = 6\sum_{j=1}^{2n-2} \frac{j}{2n-2}\left(1 - \frac{j}{2n-2}\right)\frac{1}{2n-2} = \frac{(2n-1)(2n-3)}{(2n-2)^2} \qquad (22)$$

For the dependence of $Y$ on $X$ or $Z$, given any value of $X$ or $Z$, there are two possible values of $Y$, each with $\frac{1}{2}$ probability. Conditional on the values of $X$ or $Z$, the cumulative distribution of $Y$ jumps from 0 to $\frac{1}{2}$ and then to 1.

$$\tau(X,Y)^2 = \tau(Z,Y)^2 = 6 \sum_{i=1}^{2n-2} \sum_{j=1}^{4n-4} F(j|i)^2 \frac{1}{(2n-2)(4n-4)} - 6 \sum_{j=1}^{4n-4} F(j)^2 \frac{1}{4n-4}$$

$$= \frac{6}{(2n-2)(4n-4)} \sum_{i=1}^{2n-2} \left((2i-1)\cdot\left(\frac{1}{2}\right)^2 + (2n-i-1)\cdot 1^2\right) - 6 \sum_{j=1}^{4n-4} \left(\frac{j}{4n-4}\right)^2 \frac{1}{4n-4}$$

$$= \frac{6}{(2n-2)(4n-4)}\left((2n-2)^2 \cdot \left(\frac{1}{2}\right)^2 + \frac{(2n-2)(2n-1)}{2}\cdot 1^2\right) - \frac{(4n-3)(8n-7)}{(4n-4)^2}$$

$$= \frac{1}{4} \cdot \frac{(2n-1)(2n-3)}{(2n-2)^2} \qquad (23)$$

For the dependence of $X$ on $Z$ or $Z$ on $X$, again the conditional cumulative distribution jumps from 0 to $\frac{1}{2}$ and then to 1, which is also mirror symmetric on the conditioned variable.

$$\tau(X,Z)^2 = \tau(Z,X)^2 = 6 \sum_{i=1}^{2n-2}\sum_{j=1}^{2n-2} F(j|i)^2 \frac{1}{(2n-2)^2} - 6 \sum_{j=1}^{2n-2} F(j)^2 \frac{1}{2n-2}$$

$$= \frac{6}{(2n-2)^2} \cdot 2 \sum_{i=1}^{n-1} \left( (2i-1) \cdot \left(\frac{1}{2}\right)^2 + (n-i) \cdot 1^2 \right) - 6 \sum_{j=1}^{2n-2} \left(\frac{j}{2n-2}\right)^2 \frac{1}{2n-2}$$

$$= \frac{6}{(2n-2)^2} \cdot 2 \cdot \left( (n-1)^2 \cdot \left(\frac{1}{2}\right)^2 + \frac{n(n-1)}{2} \cdot 1^2 \right) - \frac{(2n-1)(4n-3)}{(2n-2)^2}$$

$$= \frac{1}{4} \cdot \frac{n(n-2)}{(n-1)^2} \tag{24}$$

In the limit $n \to \infty$, $\tau(Y,X) = \tau(Y,Z) = 1$, which represent complete dependence or functional relationship; $\tau(X,Y) = \tau(Z,Y) = \tau(X,Z) = \tau(Z,X) = \frac{1}{2}$, which corresponds to half dependence of $Y$ on $X$, $Y$ on $Z$, $Z$ on $X$ or $X$ on $Z$. Comparing to mutual information, we notice that the new dependence measure better describes the dependence relationship between pairs from $X, Y, Z$. We remark that, if $Y$ is uniformly distributed on $[0,1]$, the continuous version of the nonsymmetric dependence measure defined in Li (2015a) gives the same result as the above $n \to \infty$ limit.

### 3. Multivariate discrete dependence measures

In Li (2015b), nonsymmetric multivariate dependence measures were constructed for continuous random variables. In the discrete case, we can also define measures that characterize the dependence of one group of discrete random variables on another group of discrete random variables.

For this purpose, we need the cumulative distribution $F(j_1 \cdots j_e | i_1 \cdots i_d)$ of one group conditional on the other and the two marginal distributions for each group, $P(i_1 \cdots i_d)$ and $P(j_1 \cdots j_e)$. The measure similar to Equation (2) is defined as

$$\tau(X,Y)^2 = 6 \sum_{i_1 \cdots i_d, j_1 \cdots j_e=1}^{m_1 \cdots m_d, n_1 \cdots n_e} [F(j_1 \cdots j_e | i_1 \cdots i_d) - F(j_1 \cdots j_e)]^2 \cdot P(i_1 \cdots i_d) \cdot P(j_1 \cdots j_e) \tag{25}$$

where $X$ and $Y$ are now vectors of random variables in each group and

$$F(j_1 \cdots j_e | i_1 \cdots i_d) = \frac{\sum_{j_1' \cdots j_e'=1}^{j_1 \cdots j_e} P(i_1 \cdots i_d, j_1' \cdots j_e')}{P(i_1 \cdots i_d)} \tag{26}$$

If $P(i_1 \cdots i_d) = 0$, then $F(j_1 \cdots j_e | i_1 \cdots i_d)$ is set to 0. Again, $\tau(X,Y) = 0$ means $X$ and $Y$ are independent

$$F(j_1 \cdots j_e | i_1 \cdots i_d) = F(j_1 \cdots j_e) \tag{27}$$

For the upper bound,

$$\tau(X,Y)^2 =$$
$$6 \sum_{i_1 \cdots i_d, j_1 \cdots j_e=1}^{m_1 \cdots m_d, n_1 \cdots n_e} F(j_1 \cdots j_e | i_1 \cdots i_d)^2 \cdot P(i_1 \cdots i_d) \cdot P(j_1 \cdots j_e) - \sum_{j_1 \cdots j_e=1}^{n_1 \cdots n_e} F(j_1 \cdots j_e)^2 \cdot P(j_1 \cdots j_e)$$

$$\leq 6 \sum_{i_1 \cdots i_d, j_1 \cdots j_e=1}^{m_1 \cdots m_d, n_1 \cdots n_e} F(j_1 \cdots j_e | i_1 \cdots i_d) \cdot P(i_1 \cdots i_d) \cdot P(j_1 \cdots j_e) - \sum_{j_1 \cdots j_e=1}^{n_1 \cdots n_e} F(j_1 \cdots j_e)^2 \cdot P(j_1 \cdots j_e)$$

$$= 6 \sum_{j_1 \cdots j_e=1}^{n_1 \cdots n_e} [F(j_1 \cdots j_e) - F(j_1 \cdots j_e)^2] \cdot P(j_1 \cdots j_e) \tag{28}$$

where we have used $0 \leq F(j_1 \cdots j_e | i_1 \cdots i_d) \leq 1$ and

$$\sum_{i_1 \cdots i_d=1}^{m_1 \cdots m_d} F(j_1 \cdots j_e | i_1 \cdots i_d) \cdot P(i_1 \cdots i_d) = F(j_1 \cdots j_e). \tag{29}$$

The equal sign holds when $F(j_1 \cdots j_e | i_1 \cdots i_d) \in \{0,1\}$. Thus, for each $j$ in $j_1, \cdots, j_e$, $F(j | i_1 \cdots i_d) = F(m_1 \cdots j \cdots m_e | i_1 \cdots i_d) \in \{0,1\}$ is non-decreasing in $j$. It jumps to 1 for a specific value of $j$ for each $i_1 \cdots i_d$, which implies that each $Y_j$ is a function of $X$. So the vector $Y$ will be function of the vector $X$. The upper bound in Equation (28) is a discrete approximation of the continuous bound expressed in terms of the Kendall distribution function in Li (2015b), Equation (48). It depends on the cumulative distribution of $Y$, but is bounded in the region $(0, \frac{3}{2})$ as $0 \leq F(j_1 \cdots j_e) - F(j_1 \cdots j_e)^2 \leq \frac{1}{4}$.

The entropy form of the dependence measure similar to Equation (5) is

$$R_\alpha(X,Y) = \frac{1}{\alpha-1} \log \left( \sum_{i_1 \cdots i_d, j_1 \cdots j_e=1}^{m_1 \cdots m_d, n_1 \cdots n_e} \left( \frac{F(j_1 \cdots j_e | i_1 \cdots i_d)}{F(j_1 \cdots j_e)} \right)^\alpha \cdot P(i_1 \cdots i_d) \cdot P(j_1 \cdots j_e) \right) \tag{30}$$

Its lower bound is 0, which corresponds to independence of $X$ and $Y$. Its upper bound is

$$\frac{1}{\alpha-1} \log \left( \sum_{j_1 \cdots j_e=1}^{n_1 \cdots n_e} F(j_1 \cdots j_e)^{1-\alpha} \cdot P(j_1 \cdots j_e) \right)$$

which holds when $Y$ is a function of $X$.

The other entropy form similar to Equation (9) is

$$\Delta_\alpha(X,Y) = \frac{1}{\alpha-1} \left( \sum_{i_1 \cdots i_d, j_1 \cdots j_e=1}^{m_1 \cdots m_d, n_1 \cdots n_e} \left( \frac{F(j_1 \cdots j_e | i_1 \cdots i_d)}{F(j_1 \cdots j_e)} \right)^\alpha \cdot P(i_1 \cdots i_d) \cdot P(j_1 \cdots j_e) - 1 \right) \tag{31}$$

In the limit $\alpha \to 1$, they both reduce to

$$R(X,Y) = \sum_{i_1 \cdots i_d, j_1 \cdots j_e=1}^{m_1 \cdots m_d, n_1 \cdots n_e} \frac{F(j_1 \cdots j_e | i_1 \cdots i_d)}{F(j_1 \cdots j_e)} \log \frac{F(j_1 \cdots j_e | i_1 \cdots i_d)}{F(j_1 \cdots j_e)} \cdot P(i_1 \cdots i_d) \cdot P(j_1 \cdots j_e) \tag{32}$$

Let $Z$ be another group of discrete random variables. If $X$ and $Z$ are independent conditional on $Y$, then we have $P(i_1 \cdots i_d, k_1 \cdots k_g | j_1 \cdots j_e) = P(i_1 \cdots i_d | j_1 \cdots j_e) \cdot P(k_1 \cdots k_g | j_1 \cdots j_e)$ and, similar to Equation (10), we have

$$P(k_1 \cdots k_g | i_1 \cdots i_d) = \sum_{j_1 \cdots j_e = 1}^{n_1 \cdots n_e} P(k_1 \cdots k_g | j_1 \cdots j_e) \cdot P(j_1 \cdots j_e | i_1 \cdots i_d) \quad (33)$$

This is the discrete form of the generalized $*$ product for copula as described in (Li, 2015b), and is also a generalized tensor product for the transition matrix of multi-dimensional Markov chain. Summing over $k_1 \cdots k_g$, we have

$$F(k_1 \cdots k_g | i_1 \cdots i_d) = \sum_{j_1 \cdots j_e = 1}^{n_1 \cdots n_e} F(k_1 \cdots k_g | j_1 \cdots j_e) \cdot P(j_1 \cdots j_e | i_1 \cdots i_d) \quad (34)$$

Similar to Equation (14), the general form of dependence measure is

$$\tau(X, Y) = \sum_{i_1 \cdots i_d, j_1 \cdots j_e = 1}^{m_1 \cdots m_d, n_1 \cdots n_e} \varphi(F(j_1 \cdots j_e | i_1 \cdots i_d) - F(j_1 \cdots j_e)) \cdot P(i_1 \cdots i_d) \cdot P(j_1 \cdots j_e)$$

$$(35)$$

for a convex function $\varphi$.

PROPOSITION 3.1 If $X, Y, Z$ form a Markov chain, then

$$\tau(X, Z) \leq \tau(Y, Z) \quad (36)$$

The proof is similar to that of Proposition 2.1 with the help of Jensen's inequality and the transition matrix in Equation (34). So the DPI condition still holds for the multivariate case.

PROPOSITION 3.2. If $f$ is a bijective transformation on $X$, then $\tau(f(X), Y) = \tau(X, Y)$.

This follows naturally as $X, f(X), Z$ form a Markov chain, which is actually the sufficient condition on $f$ as it needs not to be bijective. Note that the dependence measure is invariant under permutations on elements of the vector $X$, but is not invariant under permutations on elements of the vector $Y$.

4. **Conditional nonsymmetric dependence measure**

Conditional mutual information has also been used for causality detection in dynamic systems and time series analysis, see for example Hlaváčková-Schindler, Paluš, Vejmelka and Bhattacharya (2007). As previous discussion revealed, causality often implies complete dependence and mutual information is not a good measure of complete dependence. Therefore we will introduce a conditional nonsymmetric dependence measure which can be used in causality detection to determine true complete dependence.

Let $X, Y, Z$ be discrete random variables. We define the conditional measure as

$$\tau(X, Y|Z)^2 \equiv \sum_{k=1}^{l} \tau(X, Y|Z = z_k) \cdot P(k)$$

$$\begin{aligned}
&= \sum_{k=1}^{l}\left[6\sum_{i,j=1}^{m,n}[F(j|i,k) - F(j|k)]^2 \cdot P(i|k) \cdot P(j|k)\right] \cdot P(k) \\
&= \sum_{k=1}^{l}\left[6\sum_{i,j=1}^{m,n}\left(F(j|i,k)^2 \cdot P(i|k) \cdot P(j|k) - F(j|k)^2 \cdot P(j|k)\right)\right] \cdot P(k) \\
&\leq \sum_{k=1}^{l}\left[6\sum_{i,j=1}^{m,n}\left(F(j|i,k) \cdot P(i|k) \cdot P(j|k) - F(j|k)^2 \cdot P(j|k)\right)\right] \cdot P(k) \\
&= \sum_{k=1}^{l}\left[6\sum_{i,j=1}^{m,n}(F(j|k) - F(j|k)^2) \cdot P(j|k)\right] \cdot P(k) \quad (37)
\end{aligned}$$

Then $\tau(X,Y|Z) = 0$ if $F(j|i,k) - F(j|k) = 0$ for all $i,j,k$, which means that $X,Y$ are independent conditional on $Z$, or $X,Z,Y$ form a Markov chain. On the other hand, $\tau(X,Y|Z)$ takes maximum value when $F(j|i,k) \in \{0,1\}$ for all $i,j,k$, which implies that $Y$ is a function of $X$ and $Z$, according to discussion after Equation (28). Note that the maximum value depends on the conditional distribution $F(j|k)$, which makes it hard to judge when it reaches complete dependence in real data. If $F(j|k)$ is continuous for all $k$, then the maximum value will be the constant 1. So a rough estimate of conditional complete dependence would be if the measure is close to 1.

We remark that, in the continuous case, if the conditional distributions are also continuous, the conditional dependence measures can be again nonparametric or independent of the marginal distributions.

### 5. Conclusion

In this paper, we introduce nonsymmetric bivariate and multivariate dependence measures for discrete random variables. The measures are able to cover the two extremes of independence and complete dependence, while previous measures detect mostly divergence from independence and are unable to determine complete dependence. The new measures are nonsymmetric as complete dependence is also nonsymmetric. We use transition matrix on Markov chain to prove the DPI condition on the measures. Unlike the continuous case (Li, 2015a and 2015b), where the measures are defined through copula and thus are nonparametric, the discrete case depends on the marginal distributions and the upper bounds are also not constant. This might complicate the application of the nonsymmetric dependence measures in real situations. Finally we present a conditional nonsymmetric dependence measure, which may be used in causality detection.